\documentclass[preprint2,openany]{aastex7}

\shorttitle{Probing evolution of Long GRB properties}

\shortauthors{Khatiya et al.}

\graphicspath{{./}{figures/}}
\usepackage{graphicx}
\usepackage{tikz}
\usepackage{float}
\usepackage{amsmath}
\usepackage{subfigure}
\usepackage{xcolor}
\usepackage{soul}
\usepackage{mathtools}
\usepackage{environ}
\usepackage{multirow}
\usepackage{breqn}
\NewEnviron{myequation}{%
\begin{equation}
\scalebox{1.1}{$\BODY$}
\end{equation}}
\newtagform{smalleqnum}[\small]{\small (}{)}

\setstcolor{red}
\begin{document}

\title{Probing evolution of Long GRB properties through their cosmic formation history}

\author[0009-0002-2068-3411]{Nikita S. Khatiya}\email{nkhatiy@clemson.edu}\thanks{Email: nkhatiy@clemson.edu}
\affiliation{Department of Physics and Astronomy, Clemson University, Clemson, SC 29634, USA}

\author[0000-0003-4442-8546]{Maria Giovanna Dainotti}\email{maria.dainotti@nao.ac.jp}
\affiliation{National Astronomical Observatory of Japan, Mitaka, Tokyo 181-8588, Japan}
\affiliation{The Graduate University for Advanced Studies, SOKENDAI, Kanagawa 240-0193, Japan}
\affiliation{Space Science Institute, Boulder, CO 80301, USA}
\affiliation{Center for Astrophysics, University of Nevada, Las Vegas, NV 89154, USA}
\affiliation{Bay Environmental Institute, P.O. Box 25, Moffett Field, CA 94035, USA}

\author[0000-0003-1265-2981]{Aditya Narendra}\email{narendraaditya8@gmail.com}
\affiliation{Doctoral School of Exact and Natural Sciences, Jagiellonian University, Kraków, Poland}
\affiliation{Astronomical Observatory of Jagiellonian University, Kraków, Poland}

\author[0009-0003-8022-8151]{Dhruv S. Bal}\email{dbal@clemson.edu}
\affiliation{Department of Physics and Astronomy, Clemson University, Clemson, SC 29634, USA}

\author[0000-0003-1943-010X]{Aleksander Ł. Lenart}\email{aleksander.lenart@student.uj.edu.pl}
\affiliation{Astronomical Observatory of Jagiellonian University, Kraków, Poland}

\author[0000-0002-8028-0991]{Dieter H. Hartmann}\email{hdieter@clemson.edu}
\affiliation{Department of Physics and Astronomy, Clemson University, Clemson, SC 29634, USA}

\begin{abstract}
The astrophysics of Long GRB (LGRB) progenitors as well as possible cosmological evolution in their properties still poses many open questions. Previous studies suggest that the LGRB rate density (LGRB-RD) follows the cosmic star formation rate density (SFRD) only at high-$z$ and attribute this to the metallicity evolution of progenitor stars. 
For low $z$, opinions differ on whether the uptick in the LGRB-RD is due to a distinct class of low-luminosity GRBs or perhaps even a different progenitor subclass. 
To investigate these questions, 
we utilize data from the Neil Gehrels \emph{Swift} Observatory and ground-based observatories (redshift). To test the hypothesis that the observations can be mapped (with/without evolution) to the well-established cosmic SFRD, we consider three cases: no evolution, beaming angle evolution, and a simple power-law evolution. The comparison shows that the ‘no evolution’ case can be ruled out. Our study highlights that the beaming angle evolution or the simple power law evolution are also not sufficient to obtain a good match between the LGRB-RD and SFRD. Rather, the inclusion of multiple evolving properties of LGRBs in combination appears to be required to match the two rate densities in their entirety.
\end{abstract}

\keywords{Gamma ray bursts, Cosmology, Star Formation}

\section{Introduction} \label{sec: intro}
Gamma-ray bursts (GRBs) are transients detected at cosmological scales, with spectroscopic redshift ($z_{spec}$) observed up to $\approx$ 8.2 \citep{Tanvir2009Nature} and photometric redshift ($z_{photo}$) detected up to $\approx$ 9.4 \citep{Cucchiara2011}. In the initial stages of GRB explosion, the relativistic jets produce high-energy photons (mainly X-rays and $\gamma$-rays), often called the prompt emission phase. 
The duration corresponding to this prompt phase can be measured by burst detectors such as \emph{Swift}-Burst Alert Telescope (BAT, \cite{2005SSRv..120..143B}),  commonly called the $\mathrm{T_{90}}$ duration. The physical interpretation of $\mathrm{T_{90}}$ is unclear; thus, alternative definitions of the duration of the GRB with respect to the progenitor systems have been proposed \citep{2007ApJ...655L..25Z, 2025JHEAp..45..325Z}. Following the prompt phase, the jet decelerates and becomes capable of interacting with the circumburst medium, giving rise to emission in low-energy bands spanning almost the entire electromagnetic spectrum, often called the GRB afterglow emission phase. 

Observationally, these bursts come in two flavors: Short ($\mathrm{T_{90} < 2s }$) and Long GRBs ($\mathrm{T_{90} > 2s }$) \citep{1993ApJ...413L.101K}. 
The majority of observations indicate that the core collapse of massive stars results in LGRBs (e.g. \citet{1993ApJ...405..273W, 2003Natur.423..847H, 2003ApJ...591L..17S,  2006ARA&A..44..507W, 2013ApJ...776...98X}), whereas, the merger of BH-Neutron star (NS) or NS-NS systems results in Short GRBs (SGRBs) (e.g. \citet{1992ApJ...395L..83N, 1992ApJ...392L...9D, 2007PhR...442..166N, 2017ApJ...846L...5G, 2017ApJ...848L..14G, 2017ApJ...848L..13A}). However, the exact mechanism that drives the prompt emission and its duration is not well understood. 
Localization of LGRBs has shown that many of them occur in star-forming regions \citep{2006Natur.441..463F, 2016ApJ...817..144B, 2017MNRAS.467.1795L} while SGRBs are found in elliptical galaxies, i.e. regions with less active star formation \citep{2005Natur.437..851G}. There are exceptions to this scenario presented in the following studies: \citet{2008ApJ...677..441C, 2010ApJ...722.1946B, 2012ApJ...758...92L, 2016ApJ...821..132L}. In general, these different progenitor locations of SGRBs and LGRBs enable us to understand their connection with the SFR in the Universe. Due to the short lifetime of massive stars ($\lessapprox  40$ Myr), it is expected that LGRBs follow cosmic SFRD closely in time \citep{2008ApJ...673L.119K, 2022A&A...666A..14S}, whereas SGRBs have delayed SFRD due to the time required to undergo mergers \citep{2006A&A...453..823G, 2016A&A...594A..84G}. 

Multiple studies have extensively explored LGRB-RD and its comparison with SFRD.
A quick tabular summary of the same is provided in Table \ref{table: GRB-RD lit} of the Appendix. 
We provide a synopsis of their findings in three parts: LGRBs are SFR tracers at all $z$, and the discrepancy of LGRB-RD with respect to SFRD at high-z and low-z, respectively. Firstly, the LGRBs as SFR tracers have been studied by \citet{2006ApJ...642..371K, Pescalli:2015yva}, and \citet{2006ApJ...651..142H} till $z\sim$ 6. This match has been explained by the collapsar origin of LGRBs. 

The LGRB-RD at high $z$ is considered to be a biased tracer of star formation but with a shallower decline compared to SFRD \citep{2002ApJ...574..554L, 2004ApJ...609..935Y, 2007JCAP...07..003G, 2008ApJ...673L.119K, 2010ApJ...711..495B, 2010MNRAS.406.1944W}. These studies conclude that LGRB-RD matches well with SFRD at lower redshifts, particularly below $z$ $\sim$ 2. 
However, the comparatively higher LGRB-RD at high-$z$ has been explained by metallicity dependence and evolving initial mass function (IMF). Lastly, the low-$z$ excess in LGRB-RD compared to SFRD was shown for the first time by \citet{2015ApJS..218...13Y} for $z$ $<$ 1. 
Hints of this excess, along with low-luminosity bursts, were previously seen in \citet{2008MNRAS.388.1487L}.
\citet{2015ApJ...806...44P, 2017ApJ...850..161T, 2019MNRAS.488.5823L, 2021ApJ...908...83T, 2022MNRAS.513.1078D} found similar excess at the low-$z$ end of the LGRB-RD. 
Possible reasons to address this excess are unclear GRB classification such as contamination of low luminosity GRBs (LLGRBs), sample and instrument threshold selection effects.  

Thus, to address these different perspectives about the mismatch and understand other selection effects important in mapping the cosmic star formation to cosmic LGRB formation, we are studying the LGRB-RD and its implications with SFRD. This study uses LGRBs with $z$ estimates obtained from various space and ground-based observatories. This study advances the work of \citet{2013ApJ...774..157D} related to the luminosity function (LF) of LGRB with X-ray plateaus. We build on this study by updating the LF with a more recent sample of LGRB with X-ray plateaus.  Additionally, for the first time, we explore the LGRB-RD for this sample and its comparison to SFRD.The reason for selecting LGRBs with only X-ray plateau features is the homogeneity of the sample, the reduced scatter in correlations compared to established within the prompt emission \citep{2015ApJ...806...44P, 2024ApJ...963L..12P}, and to compare the previous LGRB-RD results for the LGRBs with optical plateau emission \citep{2024ApJ...967L..30D}. Moreover, we also derive LGRB-RD from SFRD using the Drake equation-like approach \citep{2016ApJ...823..154G} and finally, compare the LGRB-RD from the two approaches.

Recent James Webb Space Telescope (JWST) observations of $z$ $>$ 9 galaxies have shown modification in the SFRD at higher $z$ values \citep{2024ApJ...976L..16M, 2024MNRAS.533.3222D}. More observations of high-$z$ galaxies and GRBs will facilitate a better comparison between LGRB-RD and SFRD at high-$z$. In this study, we do not consider the modification in SFRD from the recent JWST observations but compare LGRB-RD with previous SFRD models in the literature (e.g., \citet{2006ApJ...651..142H, 2014ARA&A..52..415M}). 

Our analysis proceeds as follows.
$\S$~\ref{sec:sample_selection} provides details of our sample selection. 
$\S$~\ref{sec: Data Analysis} covers the methods used to study LGRB-LF and LGRB-RD. 
The main results are presented in $\S$~\ref{sec: Results} and $\S$~\ref{sec: lGRB-RD theoretical}. 
Finally, $\S$~\ref{sec: Discussion & Conclusion} gives our implications and conclusion. 
Throughout this analysis we use the cosmic concordance cosmology: $\mathrm{H_0}$ = 70 $\mathrm{km\ s^{-1}\ Mpc^{-1}}$, $\Omega_m$ = 0.3, and $\Omega_{\Lambda}$ = 0.7 \citep{2014ARA&A..52..415M}.

\section{Sample Selection} 
\label{sec:sample_selection}
Our study uses 242 LGRBs with plateaus detected in afterglow lightcurves from \emph{Swift} X-Ray telescope (XRT) \citep{2005SSRv..120..165B} and $z$ data from various ground and space-based instruments. The flat portion (slope $\sim$ 0) in afterglow lightcurves is defined as the plateau phase \citep{2013ApJ...774..157D, 2016ApJ...825L..20D}. The physical motivation of plateau emission resulting from the spin-down black hole scenario is explored in the \citet{2025arXiv250216204L} study. We use the same LGRB X-ray plateau sample as the \citet{2025A&A...698A..92N} study, and the detailed description of sample selection is provided in \citet{2025arXiv250216204L}. 
All these LGRBs are selected from March 15, 2005, to December 10, 2023 ($\sim 18$ years of data). These LGRBs have estimates of $\mathrm{T_a}$ (time at the end of plateau emission), $\mathrm{F_a}$ (its corresponding flux), $\alpha$ and $\beta$ (temporal decay index and spectral index at the end of plateau respectively), $\mathrm{T_{90}}$ and Fluence.

\section{Methodology}
\label{sec: Data Analysis}

\begin{figure}[t]
    \centering
    \includegraphics[width=\columnwidth]{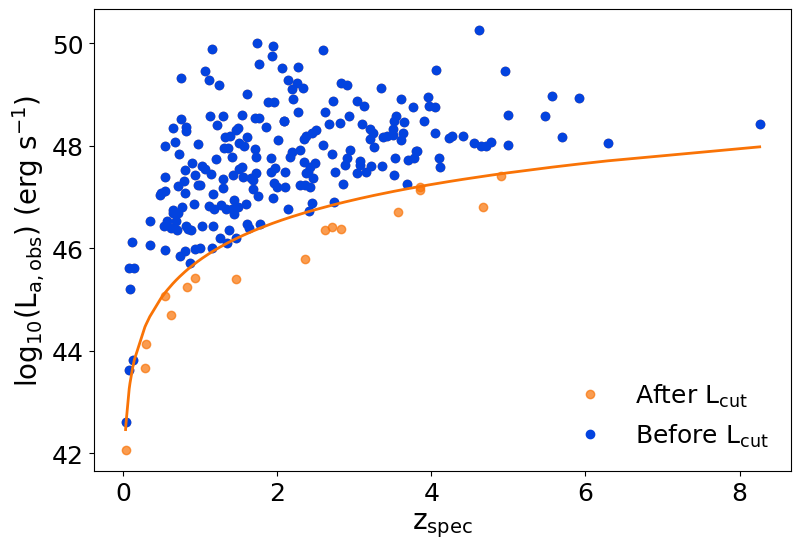}
    \caption{$\mathrm{L_{a, obs}} - z$ distribution of 242 LGRBs with X-ray plateaus. The orange solid line shows the luminosity threshold corresponding to a flux limit of $\mathrm{1.15 \times 10^{-12} \ erg \ cm^{-2} \ s^{-1}}$. Orange points highlight the 17 LGRBs cut due to this luminosity threshold and blue points correspond to LGRBs that are included further in the analysis.}
    \label{fig:Lavsz}
\end{figure}

We first calculate observed luminosity at the end of the plateau ($\mathrm{L_{a, obs}}$) and the minimum observed luminosity ($\mathrm{L_{min, obs}}$) from the available observables in the data sample given by the following expressions: 
\begin{equation}
    L_{a,obs} = 4 \pi D_L^2\frac{F_a}{(1+z)^{1-\beta}}
\end{equation}

\begin{equation}
    L_{min,obs} = 4 \pi D_L^2 F_{lim} 
\end{equation}
where, $D_L$ is the luminosity distance corresponding to $z_{obs}$. Motivated by \citet{2021MNRAS.504.4192B} and following the method highlighted in \citet{2021ApJ...914L..40D}, we do a sanity check to choose an appropriate $\mathrm{F_{lim}}$ value by understanding how the minimum flux limit can affect the LGRB-RD results. The details of this sanity check are provided in Section \ref{Limiting flux} of the Appendix. We select the most optimum case of 10\% flux limit cut, i.e., $\mathrm{\ logF_{lim} = -11.9}$. We remove 17/242 LGRBs that lie below this $\mathrm{F_{lim}}$ threshold. Figure \ref{fig:Lavsz} shows $\mathrm{L_{a, obs}} - z$ distribution along with the $\mathrm{L_{min,obs}}$ curve. 
%%%%%%%%%%%%%%%%%%%%%%%%%%%%%%%%%%%%%%%%%%%%%%
%%%%%%%%%%%%%%%%%%%%%%%%%%%%%%%%%%%%%%%%%%%%%%
To study the true nature of LGRB-RD and LGRB-LF, the intrinsic luminosity evolution needs to be corrected, which is difficult due to a significant Malmquist bias that can induce an artificial evolution. To model the luminosity evolution and correct for it in a way that it becomes independent of the Malmquist bias, we consider the non-parametric method originally developed by \citet{1971MNRAS.155...95L} for flux-limited quasar studies. One main condition of this technique is that L and 1 + $z$ should be independent variables, such that there is no luminosity evolution, to study their intrinsic luminosity and density distribution. A simple test by \citet{1992ApJ...399..345E} (EP method), a modified version of Kendall $\tau$ statistics, is employed to test the independence of L and 1 + $z$. Once the uncorrelated data set is constructed, we can then predict the intrinsic luminosity and density distributions. A detailed description of this method can be found in \citet{2013ApJ...774..157D}. 

Generally, to test the independence of two variables x and y within a data set, rank $R_i$ of $x_i$ is calculated, which is uniformly distributed between 1 and N. The expected mean E = (1/2)(N+1) and variance V = (1/12)($\mathrm{N^2}$ - 1). The Rank $\mathrm{R_i}$ is normalized with mean 0 and variance 1 and is quantified with the Kendall $\tau$ statistic given by  
\begin{equation} \label{eq: tau_calc}
    \tau = \frac{\Sigma_i (R_i - E_i)}{\sqrt{\Sigma_i V_i}}
\end{equation}

\subsection{LGRB Rate Density Analysis} \label{RDA}
The raw cumulative rate density distribution ($\sigma\mathrm{_{raw}} (< z)$) is constructed by counting the number of LGRBs below a $z$ value similar to a general cumulative distribution function. To evaluate the corrected cumulative rate density distribution ($\sigma\mathrm{_{corr} (< z)}$), modified luminosities $\mathrm{L_{a, obs} = L_{a, obs}}/(1+z)^k$ and $\mathrm{L_{min, obs} = L_{min, obs}}/(1+z)^k$ are used, where $(1+z)^k$ is the simplest functional form used to model the luminosity evolution. %%%%%%%%%%%%%%%%%%%%%%%%%%%%%%%%%%%%%%%%%%%%%%
\begin{figure}[t]
    \centering
    \includegraphics[width=\columnwidth]{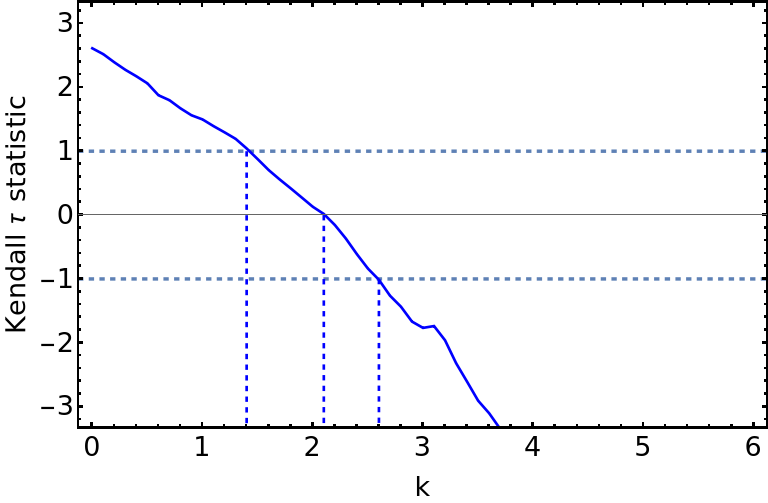}
    \caption{Kendall $\tau$ statistic - k distribution for constraining the luminosity evolution factor $(1+z)^k$. The blue dashed lines indicate k values corresponding to $\mathrm{\tau = -1, 0 \ and \ 1}$.}
    \label{fig:tau_stat}
\end{figure}
%%%%%%%%%%%%%%%%%%%%%%%%%%%%%%%%%%%%%%%%%%%%%%
We then compute the Kendall $\tau$ statistic for a range of k values (0 to 6) using the subsets called ``Associated sets (AS)" to constrain the best k-value. The AS is defined as follows: 
\begin{equation}
    J_i \equiv \left\{j: \left\{z_j < z_i\right\} \cap \left\{L_j > L_{min}(z_i) \right\} \right\}
\end{equation}

\begin{deluxetable*}{ccccc}
\tablecaption{$\sigma$ fit parameter results} \label{tab:sigma_fit_params}
\tablehead{
\colhead{Parameters} & \colhead{$\sigma_{raw}$} & \colhead{$\sigma_{raw}$} & \colhead{$\sigma_{corr}$} & \colhead{$\sigma_{corr}$} \\ \colhead{} & \colhead{$z \leq z_{cut}$} & \colhead{$z > z_{cut}$} & \colhead{$z \leq z_{cut}$} & \colhead{$z > z_{cut}$}}
%\decimalcolnumbers
\startdata
$a_{1}$ & 114.54 $\pm$ 11.41 & -200.30 $\pm$ 6.04 & 190.56 $\pm$ 33.37 & -392.27 $\pm$ 10.83 \\
$a_{2}$ & -229.83 $\pm$ 18.74 & 156.98 $\pm$ 3.64 & -373.36 $\pm$ 66.42 & 288.61 $\pm$ 7.21 \\
$a_{3}$ & 139.55 $\pm$ 9.94 & -19.26 $\pm$ 0.68 & 217.57 $\pm$ 42.31 & -32.86 $\pm$ 1.48 \\
$a_{4}$ & -20.89 $\pm$ 1.71 & 0.78 $\pm$ 0.04 & -30.34 $\pm$ 8.62 & 1.21 $\pm$ 0.09 \\
\enddata
\tablecomments{Column 1: The first four rows indicate fit coefficients ($\mathrm{a_1, a_2, a_3, a_4}$) that describe the third-order polynomial in the piecewise function as shown in equation \ref{sigma_fit}. Columns 2 and 3: $\sigma\mathrm{_{raw}}$ parameters before and after the $z_{cut}$ respectively, Columns 4 and 5: $\sigma\mathrm{_{corr}}$ parameters before and after the $z_{cut}$ respectively.}
\vspace*{-\baselineskip}
\end{deluxetable*}

For every burst $i$ with coordinates ($z_i, \mathrm{L_i}$), a data subset is constructed such that all bursts with $\mathrm{L_j > L_{min}}(z_i)$ and $z_j < z_i$ are selected. The index j represents bursts within the associated set i. $\mathrm{L_{min, i}}$ is the minimum luminosity corresponding to the $z_i$ value determined from the luminosity threshold curve. The rank $\mathrm{R_i}$ of $z_i$ is computed as follows and is uniformly distributed between 1 and $\mathrm{M_j}$. The total number of bursts in an associated set $\mathrm{J_i}$ is represented by $\mathrm{M_j}$. 

\begin{equation}
    R_i \equiv \left\{j \in J_i: z_j < z_i \right\}.
\end{equation}

Using these ranks $\mathrm{R_i}$ as per the associated sets, its corresponding Kendall $\tau$ statistic is calculated using Equation \eqref{eq: tau_calc}. The $\tau$-k distribution is shown in Figure \ref{fig:tau_stat}. The best k-value of $2.1^{+0.5}_{-0.7}$ (3$\sigma$ from the no evolution case i.e. k = 0) is selected for the $\tau$ = 0 case. The 1$\sigma$ errors in k are found from $\tau$ = -1 and $\tau$ = 1 cases. For the estimated k-value, the deevolved/modified luminosities are first calculated and then used to obtain the $\sigma\mathrm{_{corr} (< z)}$ distribution using the following expression: 
\begin{equation}
    \sigma_{corr}(< z_i) = \displaystyle \prod_{j = 2}^{i} \left[1 + \frac{1}{M_j}\right].
\end{equation}

\begin{figure*}[ht]
    \centering
    \includegraphics[width=\columnwidth]{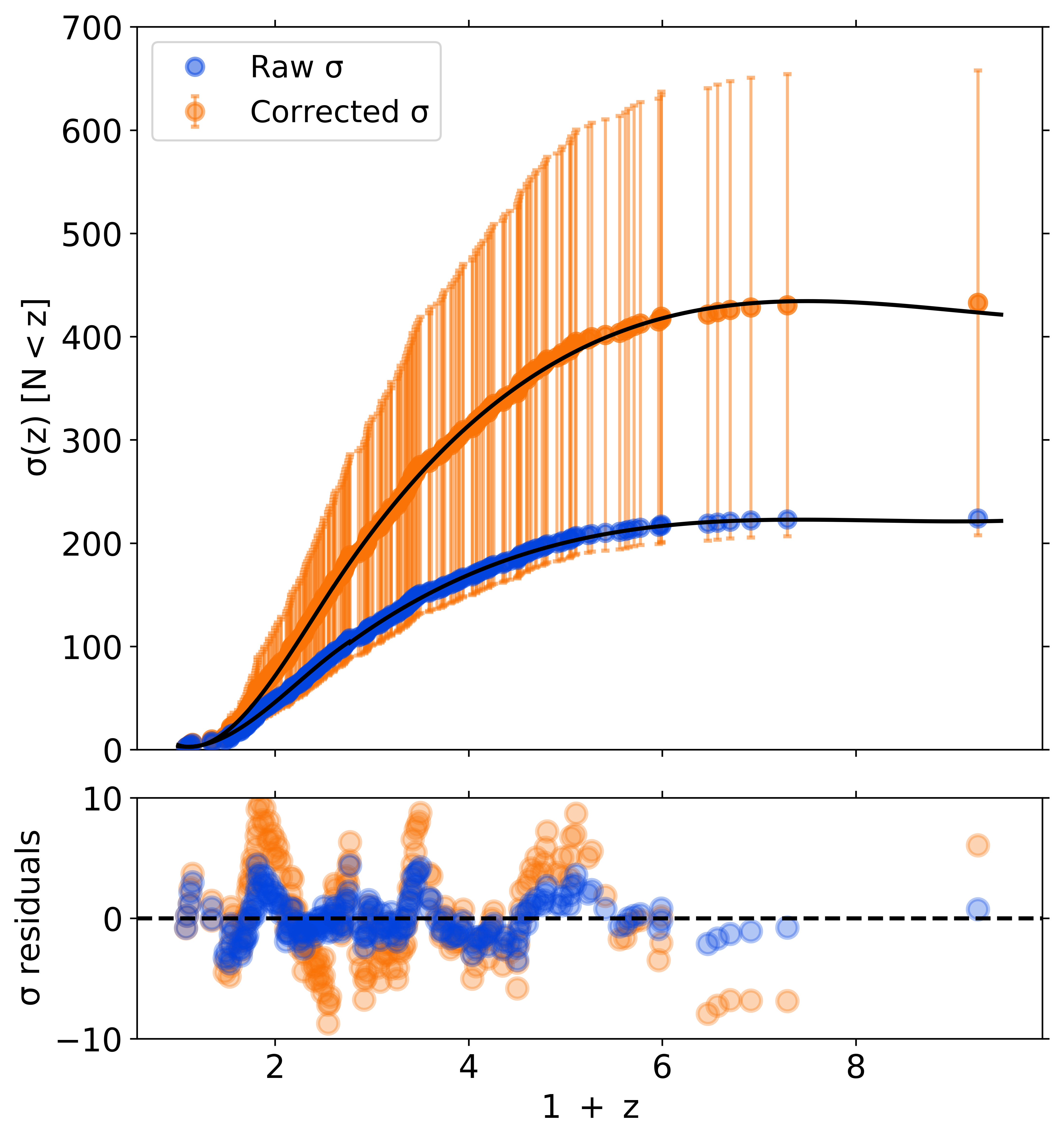}
    \includegraphics[width=\columnwidth]{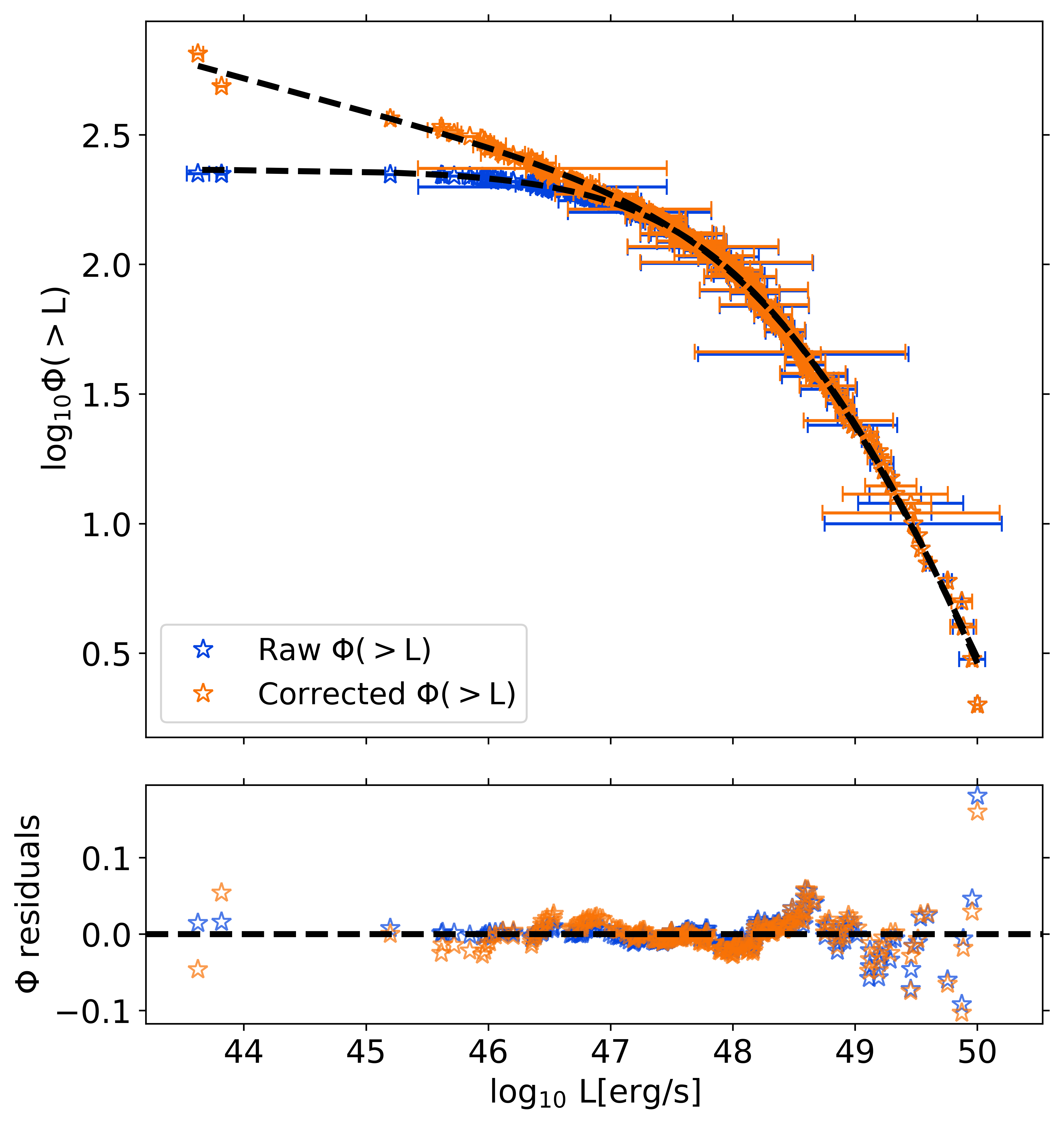} 
    \caption{Left panel: Cumulative LGRB density distribution for the raw (blue) and corrected (orange) cases. The black solid line corresponds to the piecewise functional fit used to calculate the LGRB-RD later. Right panel: log-log distribution of $\mathrm{\Phi(L)}$ and $\mathrm{L [erg \ s^{-1}]}$. The blue and orange open stars show the raw and corrected $\Phi(>L)$ points, respectively. The black dashed lines correspond to the best-fit functions described in equation \ref{LF_fit_func}. Residuals for raw and corrected $\sigma(z)$, and raw and corrected $\Phi(>L)$ cases are shown in the bottom panel.}
    \label{fig:Fitted_sigma_LF}
\end{figure*}

The left panel of Figure \ref{fig:Fitted_sigma_LF} shows the $\sigma\mathrm{_{raw}}$ and $\sigma\mathrm{_{corr}}$ distributions. The cumulative raw and corrected LGRB rate density is then best fitted with a piecewise function. The functional form of this function is: 
\begin{equation}
    \sigma_{fit} = \begin{cases}
        a_1 + a_2z + a_3z^2 + a_4z^3, z \leq z_{cut} \\

        a_1 + a_2z + a_3z^2 + a_4z^3, z > z_{cut}
    \end{cases}
    \label{sigma_fit}
\end{equation}

$z_{cut}$ is considered as 2.77 and 2.6 in the $\sigma\mathrm{_{fit}}$ function to best describe the $\sigma\mathrm{_{raw}}$ and $\sigma\mathrm{_{corr}}$ case respectively. The fit parameters are presented in Table \ref{tab:sigma_fit_params}. The fitting is done using the Levenberg-Marquardt algorithm in Mathematica. To fit the $\sigma\mathrm{_{corr}}$ data points, the inverse of the uncertainties squared are the weights that have been taken into account. Since the data points and errors in $\sigma(<z)$ are correlated with each other, $\chi\mathrm{_{red}^2}$ and $\mathrm{R^2}$ are unreliable fit metrics to determine goodness of fit. Thus, we select the best fit as long as the fit parameters are well-constrained and the fit looks visually appropriate. This best fit is further supported by the residual analysis illustrated in the bottom left panel of Figure \ref{fig:Fitted_sigma_LF}. 

%%%%%%%%%%%%%%%%%%%%%%%%%%%%%%%%%%%%%%%%%%%%%%

\subsection{Luminosity Function} \label{LF}
The raw cumulative LF ($\mathrm{\Phi_{raw} (> L)}$) is constructed by counting the number of LGRBs above a $\mathrm{L_{a, obs}}$ value similar to a cumulative distribution function. It is shown in Figure \ref{fig:Fitted_sigma_LF} below. To evaluate the corrected cumulative LF ($\mathrm{\Phi_{corr} (> L)}$), similar associated sets are generated but with a different condition as described below: 

\begin{equation}
    J_i \equiv \left\{j: \left\{z_j < z_{max}(L_i)\right\} \cap \left\{L_j > L_i\right\} \right\}
\end{equation}
Here, $z_{max}(\mathrm{L_i})$ is the maximum redshift corresponding to the $\mathrm{L_i}$ value determined by the luminosity threshold curve. The corresponding rank $\mathrm{R_i}$ is defined as 
\begin{equation}
    R_i \equiv \left\{j \in J_i: L_j > L_i \right\}
\end{equation} Also, $\mathrm{E_j}$ = (1/2)($\mathrm{N_j}$ + 1) and $\mathrm{V_j}$ = (1/12)($\mathrm{N_j^2 - 1}$), where $\mathrm{N_j}$ is the total number of bursts in an associated set $\mathrm{J_i}$. For the best k-value, the $\mathrm{\Phi_{corr} (> L_i)}$ is obtained in the following manner: 

\begin{equation}
    \Phi_{corr}(> L_i) = \displaystyle \prod_{j = 2}^{i} \left[1 + \frac{1}{N_j}\right]
\end{equation}

This $\mathrm {\Phi_{raw} (> L)}$ and $\mathrm{\Phi_{corr}(> L)}$ are fitted with a smoothly broken power law (SBPL) by accounting for errors along the L axis. The functional form of SBPL is given as
\begin{equation}
    log \ \Phi_{fit}(L) = \mathrm{log \ c} - \mathrm{\frac{log [L^{a.w} L_b^{-a.w} + L^{b.w} L_b^{-b.w}]}{w} }
    \label{LF_fit_func}
\end{equation}
The fitting is done using the orthogonal distance regression method to account for the errors in L data using Python. The best-fit parameters for $\mathrm{ \Phi_{raw} (> L)}$ and $\mathrm{ \Phi_{corr} (> L)}$ are presented in Table \ref{tab:phi_fit_params}. Since the data points and errors in $\mathrm{ \Phi (> L)}$ are correlated with each other, $\chi\mathrm{_{red}^2}$ and $\mathrm{R^2}$ are unreliable fit metrics to determine goodness of fit. Thus, we select the best fit as long as the fit parameters are well-constrained and the fit looks visually appropriate. The best-fit function is further selected based on the residual analysis method as shown in the bottom right panel of Figure \ref{fig:Fitted_sigma_LF}.  

\begin{deluxetable*}{ccc}
\tablecaption{$\Phi$ fit parameter results} \label{tab:phi_fit_params}
\tablehead{
\colhead{Parameters} & \colhead{$\Phi_{raw}$} & \colhead{$\Phi_{corr}$}} 
%\decimalcolnumbers
\startdata
$L_b$ & 48.59 $\pm$ 0.09 & 48.94 $\pm$ 0.15 \\
$a$ & 140.67 $\pm$ 7.84 & 153.38 $\pm$ 12.37 \\
$b$ & -1.86 $\pm$ 0.93 & 12.73 $\pm$ 0.83 \\
$c$ & 266.86 $\pm$ 20.25 & 134.87 $\pm$ 6.30 \\
$w$ & 0.39 $\pm$ 0.04 & 0.42 $\pm$ 0.06 \\
\enddata
\tablecomments{Column 1: The first five rows indicate fit coefficients ($\mathrm{L_b, a, b, c, w}$) that describe the smoothly broken power law as shown in equation \ref{LF_fit_func}. Columns 2 and 3: $\Phi\mathrm{_{raw}}$ and $\Phi\mathrm{_{corr}}$ parameters respectively.}
\vspace*{-\baselineskip}
\end{deluxetable*}

\section{Results} \label{sec: Results}
\subsection{Raw and Corrected LGRB-RD} \label{sec:Raw and Corrected GRB-DR}
The differential raw and corrected LGRB rate density ($\rho(z)$) is calculated using the derivative of the fit function ($\sigma\mathrm{_{fit}}$) as described in equation \ref{rho_z}.

\begin{equation}
    \rho(z) = \frac{d\sigma}{dz} (1+z) \left(\frac{dV}{dz}\right)^{-1}
    \label{rho_z}
\end{equation}

%%%%%%%%%%%%%%%%%%%%%%%%%%%%%%%%%%%%%%%%%%%%%%

%%%%%%%%%%%%%%%%%%%%%%%%%%%%%%%%%%%%%%%%%%%%%%
\begin{figure*}
    \begin{center}
    \includegraphics[width=0.9\textwidth]{
    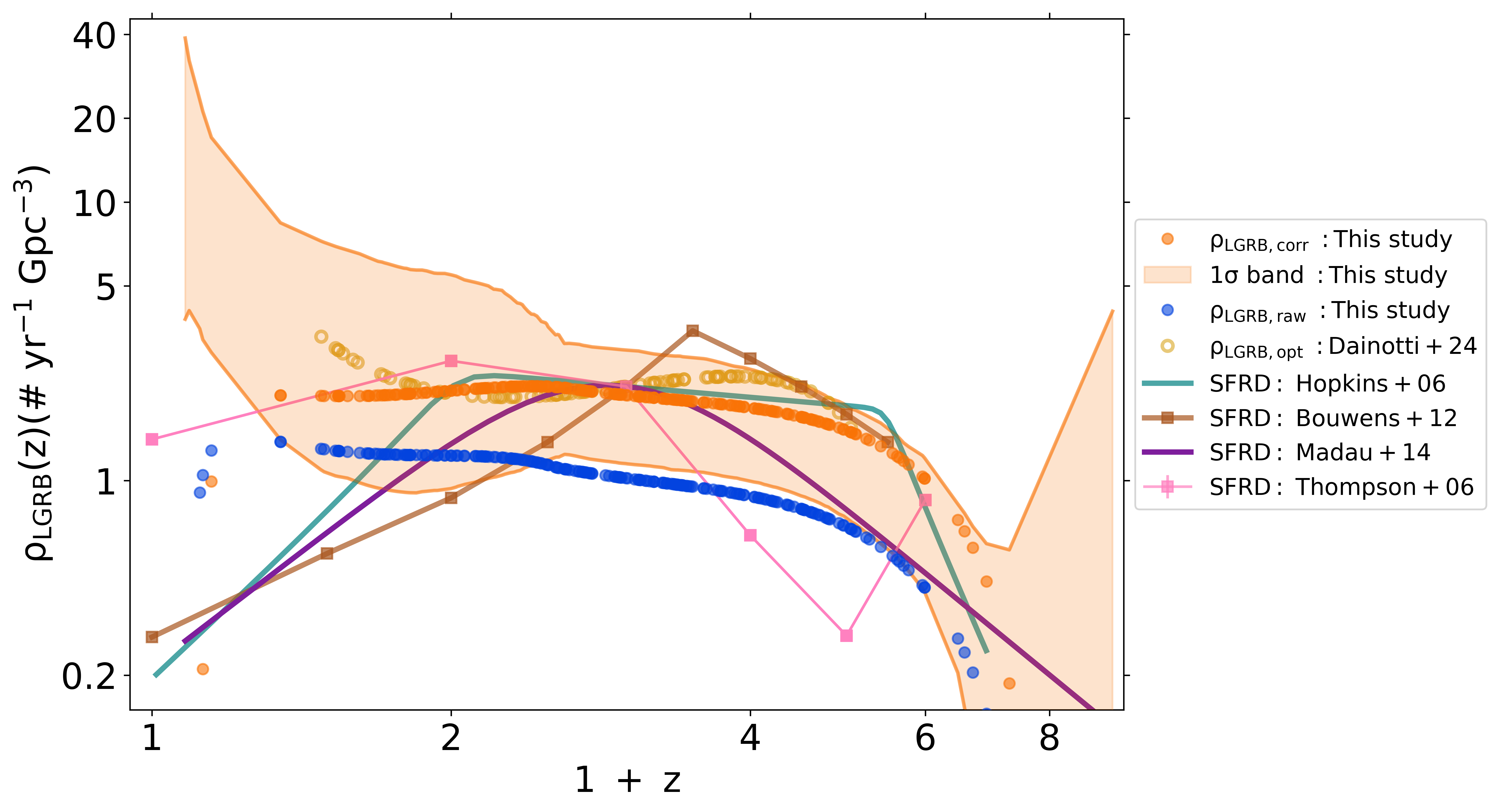}
    \caption{LGRB-RD $(\rho \ \mathrm{[yr^{-1} \ Gpc^{-3}]}) - (1+z)$ distribution for raw (solid blue points) and corrected (solid orange points) computed from our X-ray plateau sample. The orange-colored band represents the 1 $\sigma$ error band for the $\rho\mathrm{_{LGRB,corr}}$ points. Corrected LGRB-RD for optical plateau sample from \citet{2024ApJ...967L..30D} is denoted with yellow open circles. Various SFRD models from the literature are also included in this plot for comparison.}
    \label{fig:DRE}
    \end{center}
\end{figure*}
%%%%%%%%%%%%%%%%%%%%%%%%%%%%%%%%%%%%%%%%%%%%%%

This $\rho(z)$ is computed for LGRBs with X-ray plateaus only. To capture the true LGRB-RD, we need to extend our results to all LGRBs (i.e., cases when the X-ray plateau is absent or no redshift estimates are available) and correct for any LGRBs excluded during the analysis. There are two factors used to calculate true LGRB-RD, as shown in equation \ref{eq: rho_lGRB_z}. This equation is motivated by the results in \citet{2024ApJ...967L..30D}. 

\begin{equation}
    \rho_{LGRB}(z) = \frac{F_1*F_2* \rho(z)}{Years*Uptime*FOV}
    \label{eq: rho_lGRB_z}
\end{equation}

where Factors $\mathrm{F_1}$ and $\mathrm{F_2}$ are detailed as follows. $\mathrm{F_1}$ accounts for LGRBs that are excluded from the analysis. Out of 242 LGRBs in the sample, 17 LGRBs are excluded due to being below the $\mathrm{F_{lim}}$ threshold, and 2 LGRBs are dropped for AS number equal to zero. So, the analysis is carried out for 223 GRBs, resulting in $\mathrm{F_1}$ = 242/223. $\mathrm{F_2}$ accounts for LGRBs that do not have redshift estimates or X-ray plateaus, and are missed by \emph{Swift} due to Field of View (FOV), Uptime and Years of operation factors. \emph{Swift}-BAT reports the detection of 1420 LGRBs until December 10, 2023. This gives $\mathrm{F_2}$ = 1420/242. Also, the instrument uptime of 75.8\% \citep{2016ApJ...818..110B}, 18.75 years of operation (March 15, 2005 - December 10, 2023), and field of view = 0.17 ($\sim 2/12.57$ steradians) \citep{2014ApJ...783...24L} factors should be properly accounted for in the calculation of $\rho(z)$. The final $\rho\mathrm{_{raw}}(z)$ and $\rho\mathrm{_{corr}}(z)$ after accounting for all the above factors are shown in Figure \ref{fig:DRE}. 

The orange 1$\sigma$ error band in Figure \ref{fig:DRE} is constructed using the Bootstrap method for 1000 samples. We find the best distribution that describes the $z$, $\beta$ and $\mathrm{log F_a}$ parameters of the original sample and then draw 242 random values from each distribution. The 242 simulated values are drawn to match the original sample size of the analysis. The $\mathrm{L_{a, obs}}$ and $\mathrm{L_{min, obs}}$ for each simulated GRB is computed for $\mathrm{log F_{lim} = -11.9}$ case. Depending on this $\mathrm{F_{lim}}$ value, GRBs are removed and the similar steps are followed from calculating the cumulative redshift distribution ($\sigma(>z)$) to $\rho\mathrm{_{corr}}$. This step is repeated 1000 times to find a $\rho\mathrm{_{corr}}$ distribution. The 1$\sigma$ error band is then computed from the $\rho\mathrm{_{corr}}$ distribution. We can see that the error band misses the $\rho\mathrm{_{corr}}$ data points at the low z end because the $\sigma(z)$ fits are sensitive. The associated sets at the low-z and high-z ends have a smaller number of GRBs, making it difficult to constrain the fits and, thus, the $\rho\mathrm{_{corr}}$ at those extremes. Therefore, one should remain cautious when interpreting the results at both ends.  

\subsection{Raw and Corrected LGRB-LF} \label{sec:Raw and Corrected GRB-LF}

As we can see from the right panel of Figure \ref{fig:Fitted_sigma_LF}, the $\mathrm{ \Phi_{raw} (> L)}$ and $\mathrm{ \Phi_{corr} (> L)}$ are denoted with blue and orange stars respectively. It is evident that for the low luminosity tail, we have a difference between the two distributions for luminosities below $10^{47} \ \mathrm{erg \ s^{-1}}$ due to the correction for evolution. Thus, if we do not correct for evolution then we may not see a difference between the two distributions for low-luminosity and high luminosity GRBs. For L $\ge$ $10^{47} \ \mathrm{erg \ s^{-1}}$, we appreciate no substantial difference. In principle, one can take a sample which tends to be complete at high luminosity \citep{2014MNRAS.442.2342D} and  devoid of any luminosity evolution effects, but in such a case we will have a missing population of GRBs at low luminosity. It is remarkable to observe that the residuals between the fitted functions and the data shown in the bottom right panel of Figure \ref{fig:Fitted_sigma_LF} shows a very small deviation, maximum at the level of 20\%, for luminosities $\ge 10^{50} \ \mathrm{erg \ s^{-1}}$, thus showing that our fitting procedure is well suited to represent the observational data.

\section{Does $\rho\mathrm{_{corr}}$ map the $\rho\mathrm{_{theor}}$?} \label{sec: lGRB-RD theoretical}
In this study, we compare the $\rho\mathrm{_{corr}}$ computed from the X-ray plateau sample of LGRBs with the theoretical LGRB-RD $(\rho\mathrm{_{theor}})$. The $\rho\mathrm{_{theor}}$ is computed from SFRD using the ``Drake equation" approach followed in \citet{2016ApJ...823..154G} and can be expressed as:
\begin{multline}
    \rho_{theor}(z) = \rho_{*}(z) \times f_{\mathrm{*/ccSN}} \times f_{\mathrm{ccSN/SN1bc}} \times \\ f_{\mathrm{SN1bc/SN1c}} \times f_{\mathrm{beam}}\times f_{\mathrm{z}},
    \label{eq: Drake_eqn}
    \end{multline}
where $\rho\mathrm{_{*}}(z)$ is the SFRD taken from \citet{2014ARA&A..52..415M} in units of $\mathrm{M_{\odot} \ yr^{-1} \ Mpc^{-3}}$ and its functional form is given as: 
\begin{equation}
    \rho_{*}(z) = 0.015 \frac{ (1+z)^{2.7}}{1 + (\frac{1 + z}{2.9})^{5.6}}
\end{equation}
The $f_{\mathrm{*/ccSN}}$ is calculated using Salpeter IMF \citep{1955ApJ...121..161S} for a mass range of 12 to 100 $\mathrm{M_{\odot}}$ as given in equation \ref{IMF: Star to ccSN}. 

\begin{figure*}
\begin{interactive}{js}{grbr_plotly_errband.zip}
\includegraphics[width=0.9\textwidth]{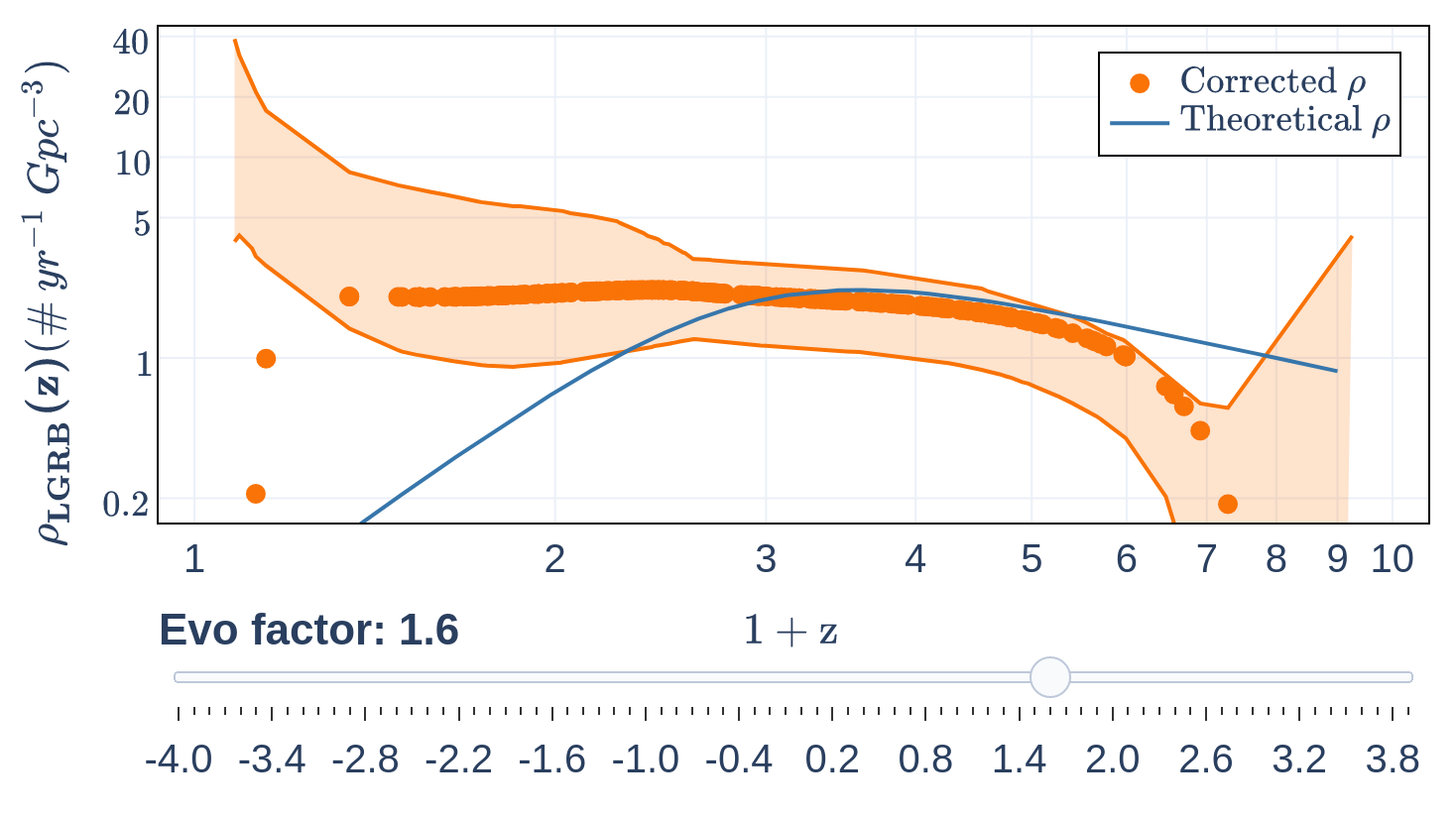}
\end{interactive}
\caption{Example figure of $\rho\mathrm{_{LGRB}(z)}$ in $1+z$ space. The orange-filled circles and the corresponding orange band represent the $\rho\mathrm{_{corr}}$ computed from Equation \ref{eq: rho_lGRB_z} and its 1$\sigma$ band. The blue solid curve is $\rho\mathrm{_{theor}}$ derived from Equation \ref{eq: Drake_eqn} and multiplied with $(1+z)^{\delta}$.  This example figure is one case of the interactive figure with $\delta$ = 1.6. The interactive figure, available in html format, has a slider for the evolution factor ($\delta$) at the bottom of the figure. The slider will help in visualizing the variation in $\rho\mathrm{_{theor}}$ for a changing $\delta$ value.}
\label{fig:interactive}
\end{figure*}

\begin{equation} 
\label{IMF: Star to ccSN}
    f_{\mathrm{*/ccSN}} = \frac{\int_{M_{min}}^{M_{max}} M^{-2.35} dM}{\int_{0.1}^{100} M M^{-2.35} dM} = 4.2 \times 10^{-3} M_{\odot}^{-1}
\end{equation}

The $f_{\mathrm{ccSN/SN1bc}}$ factor is taken from \citet{2020ApJ...904...35P}, where it is calculated as $\mathrm{f_{ccSN/H-poor}* f_{H-poor/SN1bc}} = 27.8\% * 58.6\% = 16.3\%$. The $f_{\mathrm{SN1bc/SN1c}}$ is the fraction of SN1c out of SN1bc types and is considered as 69\% from \citet{2016ApJ...823..154G}. The $f_{\mathrm{beam}}$ is considered to be about 1\% \citep{2022ApJ...929..111F}, giving us the fraction of GRBs that are beamed towards Earth for our detection purpose; $f_{\mathrm{z}}$ accounts for the LGRBs hosted in low as well as high metallicity environments in the host galaxies. This factor is explained in detail in \citet{2016ApJ...823..154G}. Initially, no redshift evolution factor is considered to compare the $\rho\mathrm{_{theor}}$ with the $\rho\mathrm{_{corr}}$. In that case, the peak and slopes of both the rate densities are incompatible with each other. To check the compatibility, we introduce a $z$-evolution term that could arise from factors in equation \ref{eq: Drake_eqn}. We consider the beaming evolution factor of $(1+z)^{1.6}$ as studied in \citet{2020MNRAS.498.5041L} to understand its impact on $\rho\mathrm{_{theor}}$. This evolution factor flattens the high-$z$ slope to match better with $\rho\mathrm{_{corr}}(z)$, however, the low-$z$ end is not explained. This indicates a requirement for another evolution factor that will be needed to describe better the $\rho\mathrm{_{corr}}(z)$ on both the low-$z$ and high-$z$ ends. We design an interactive plot showing how a simple power law evolution factor of $(1+z)^{\delta}$ behaves for $\delta \in (-4, 4)$. The interactive plot shows that the negative values of $\delta$ can only explain the low-$z$ end, while the positive $\delta$ values can only explain the high-$z$ end. However, a single $(1+z)^{\delta}$ factor in $\rho\mathrm{_{theor}}$ cannot fully map the $\rho\mathrm{_{corr}}(z)$, indicating a need for a more involved evolution factor. The $(1+z)^{1.6}$ and $(1+z)^{\delta}$ cases can be visualized with Figure \ref{fig:interactive}.  
\section{Discussion and Conclusion} \label{sec: Discussion & Conclusion}
The state-of-the-art LGRB-RD view tells us that LGRBs are not necessarily direct tracers of cosmic SFRD, with few exceptions. 
Exceptions like \citet{2006ApJ...642..371K} and \citet{Pescalli:2015yva} using BATSE and BAT6 samples and deep sky surveys, as well as in \citet{2006ApJ...651..142H} for redshifts up to 6, have shown that LGRBs are direct tracers of SFR. 
The authors explain that this is because core-collapse supernovae born from massive stars, are considered progenitors of LGRBs. 

However, majority of the literature indicates that LGRBs are biased tracers of SFR. At high-z, the mismatch can be interpreted by the metallicity dependence and evolving IMF. The metal-poor massive stars do not lose their outer envelope easily due to stellar winds \citep{2006ARA&A..44..507W,2005A&A...442..587V, 2022A&A...666A..14S}], leading to an increase in their angular momentum and eventually, a higher chance of GRB formation. 
This effect is seen as prevalent at high-$z$ due to more metal-poor stars in this redshift range. 
Another effect to explain higher LGRB-RD at high-$z$ is evolving IMF \citep{1998MNRAS.301..569L, 2008ApJ...674...29V}, i.e., IMF at high-$z$ could flatten out, leading to more massive star formation, thereby increasing the chances of LGRB formation as well.
Other possible explanations are an increased rate of closed binaries that can produce GRBs or a large population of GRB host galaxies at high-$z$ \citep{2008ApJ...673L.119K}. 

The discrepancy of LGRB-RD and SFRD at low-$z$ could be a combination of several effects discussed in the subsequent paragraph. 
Usually, GRBs are classified empirically \citep{2009ApJ...703.1696Z, 2017Ap&SS.362...70K, 2020MNRAS.492.1919M}, which might not always be the best method to distinguish sources of different origins, thus leading to sample contamination. A physically justified approach based on modeling all observed features of a given source would help to avoid contamination from SNe unassociated GRBs or merger-origin LGRBs  \citep{2024ApJ...963L..12P}.
 
Sample incompleteness, i.e., a higher number of low-$z$ GRBs in the sample, can lead to an inherent bias, leading to the low-$z$ excess seen in LGRB-RD.
Furthermore, inconsistencies in redshift measurement, GRB localization, and challenges in the rapid afterglow follow-up with optical/IR telescopes could also result in the excess. 
Differences in the progenitor system of LLGRBs, a subset of GRBs that have different physical properties compared to usual ones \citep{2011ApJ...739L..55B, 2015ApJ...807..172N}, and high-luminosity GRBs (HLGRBs) could also cause contamination from LLGRBs in the low-$z$ end. 
\citet{2022MNRAS.513.1078D} have shown the existence of a Gaussian-like component of LGRB-RD for $z<$ 1, irrespective of the completeness of the sample, a hint of the physical origin of this low-$z$ excess in the LGRB-RD.

To discern the cause of these differences between LGRB-RD and SFRD $z$-evolution, we explore a similar idea in our study using LGRBs with X-ray plateaus and $z$-estimates. We use the non-parametric EP method by correcting the LGRB-RD for intrinsic luminosity evolution. This $\rho\mathrm{_{corr}}(z)$ is compared with the $\rho\mathrm{_{theor}}(z)$ to understand the possible implications of the selection bias. 

The $\rho\mathrm{_{corr}}(z)$, when directly compared with multiple SFRD models in the literature, shows quite different behavior in the high-$z$ and low-$z$ tails. $\rho\mathrm{_{corr}}(z)$ remains almost flat for most of the $z$-range with a subtle peak around (1+$z$) $\sim$ 2.5. The LGRB rate is higher than SFR at the low-$z$ end for almost all models in the literature. In the high-$z$ regime, the Hopkins+06 (teal curve) in Figure \ref{fig:DRE} partially matches the $\rho_{corr}(z)$, but there is a mismatch with the rest of the SFRD curves. Also, the SFRD peak is at a slightly higher $z$-value than the $\rho\mathrm{_{corr}}(z)$ curve, indicating a $z$ evolution of LGRB progenitors i.e. the evolution of massive stars. We also compare the $\rho\mathrm{_{corr}}(z)$ curve with the LGRB-RD calculated from 97 LGRBs with an optical plateau sample \citep{2024ApJ...967L..30D}. Since both curves compare the LGRB-RD, their values should match. Due to the smaller sample size, i.e., $\sim$ 2.3 times less than our sample, and possibly due to different progenitors, the LGRB-RD of the optical sample shows a higher value at the low-$z$ end and slightly higher at the high-$z$ end in contrast to our study. 

Upon assessing the $\rho\mathrm{_{corr}}(z)$ with the LGRB-RD cited in the literature, we see that our study also indicates an excess at lower $z$, in agreement with the \citet{2007JCAP...07..003G, 2008MNRAS.388.1487L, 2015ApJ...806...44P, 2015ApJS..218...13Y, 2019MNRAS.488.5823L, 2021ApJ...908...83T, 2022MNRAS.513.1078D} and \citet{2024ApJ...963L..12P}. However, this low-$z$ excess starts to decrease at $z \sim$ 0.1, which can be attributed to the low number of LGRBs within the corresponding comoving volume. This analysis also exhibits an excess at the high-$z$ portion of $\rho\mathrm{_{corr}}(z)$ compared to previously studied LGRB-RD, which most likely could be due to the high probability of forming massive stars from the low-metallicity environments at higher $z$ \citep{2022ApJ...929..111F, 2022ApJ...932...10G}. 

To learn more about the properties of the LGRB progenitor system, we compare $\rho\mathrm{_{theor}}(z)$ calculated from cosmic SFRD using the Drake equation method with $\rho\mathrm{_{corr}}(z)$. This mapping indicates that $\rho\mathrm{_{theor}}(z)$ is incompatible with $\rho\mathrm{_{corr}}(z)$ and requires a modification in $\rho\mathrm{_{theor}}(z)$ to capture the idea of the parent population of LGRB fully. The effect of $(1+z)^{1.6}$ on beaming evolution studied by \citet{2020MNRAS.498.5041L} flattens the high-$z$ end of $\rho\mathrm{_{theor}}(z)$, however, the low-$z$ $\rho\mathrm{_{theor}}(z)$ drops significantly. This tells us that the simple evolution of the jet opening angle ($f_b \propto \theta^{-2}$) is not sufficient to map the cosmic LGRB formation rate to the cosmic star formation rate. Multiple selection biases need to be carefully accounted for to reveal the properties of these progenitors. To fathom the nature of the selection bias, we design an interactive plot with $(1+z)^{\delta}$ evolution term in $\rho\mathrm{_{theor}}(z)$. This interactive plot shows that either low-$z$ (negative $\delta$ values) or high-$z$ (positive $\delta$ values) end of the $\rho\mathrm{_{theor}}(z)$ curve could be modified to match the $\rho\mathrm{_{corr}}(z)$ partially. A general $(1+z)^{\delta}$ evolution has difficulties in explaining the $\rho\mathrm{_{corr}}(z)$ and hence, we desire a more elaborate redshift evolution function that takes into account metallicity bias or even evolution in progenitor SN systems. In addition, this mismatch could still exist due to LLGRB contamination or GRB driven by mergers in the current study, since the canonical split $\mathrm{T_{90}} = 2s$ is considered in the LGRB classification. 
Indeed, in \cite{2024ApJ...963L..12P}, a possible explanation is that the low-$z$ GRBs are driven by a merger scenario instead of a collapsar one. Thus, more data, possibly determined with redshift inference via machine learning, see \citep{2025A&A...698A..92N}, will help us further address these challenging issues.

The derived $\rho\mathrm{_{corr}}(z)$ is a powerful tool, not only limited to the investigation of LGRB progenitors from a star formation history, but also serves as a complementary tool to probe the stellar mass BH population rate formed from LGRBs. This interesting perspective has been presented in \citet{2018A&A...610A..58A} and \citet{2022ApJ...933...17A}. Additionally, it gives insight about the LGRB hosts, i.e., stellar or binary evolution properties \citep{2011MNRAS.416.2760C, Lloyd-Ronning:2024zgs}, their metallicity evolution \citep{2017ApJ...834..170G}, testing cosmological parameters \citep{2022MNRAS.514.1828D}, and probing the nature of high-$z$ galaxy populations \citep{2023A&A...671A..84S}.

%\begin{acknowledgments}
\section*{Acknowledgements}
N.S.K. acknowledges Clemson University for the license support received for the Mathematica software. M.G.D. acknowledges the support of the JSPS Grant-in-Aid for Scientific Research (KAKENHI) (A), Grant Number JP25H00675. We thank Biagio De Simone for his insights during the initial stages of the project. Additionally, we would like to thank the anonymous referee for their helpful suggestions.
%\end{acknowledgments}
\software{This work benefited from the following software: \textsc{Mathematica} \citep{Mathematica},
\textsc{Scipy} \citep{2020SciPy-NMeth}, \textsc{Matplotlib} \citep{Hunter:2007}.}

\bibliography{GRB}{}
\bibliographystyle{aasjournal}
\let\pagebreak\relax
\section{Appendix}
\subsection{LGRB-RD Summary} Table \ref{table: GRB-RD lit} summarizes the literature on how LGRB-RD compares to cosmic SFRD in different redshift ranges, whether the analysis has accounted for luminosity evolution or not, and the proposed solution for the discrepancy between the two rate densities.
\startlongtable
\begin{deluxetable*}{ccccc}
\tablecaption{Summary of all GRB-RD studies done in the literature}
\label{table: GRB-RD lit}
\tablehead{ 
\colhead{Reference}  & \colhead{Matching} &  \colhead{Not matching} & \colhead{Account for} & \colhead{Solution proposed} \\ \colhead{} & \colhead{SFR at} & \colhead{SFR at} & \colhead{L-evo?} & \colhead{for discrepancy}}
\startdata 
\citet{2002ApJ...574..554L} & $z<2$ & $z>2$ & Yes, $(1+z)^{1.0}$ & \multirow{12}{8em} 
    {Z decreases with an increase in $z$, resulting in a higher GRB-R due to the preference of
    low-Z environment.} 
\\
\citet{2004ApJ...609..935Y} & $z<2$ & $z>2$ & Yes, $(1+z)^{2.6}$& \\
\citet{2007JCAP...07..003G} & $-$ &Paucity:$1<z<2$ & $-$ &\\ 
\citet{2008ApJ...673L.119K} &$-$ & $z<2$ for LLGRB excess& No,might cause &\\ 
 & & $z>3$ for HLGRB excess & excess at low-$z$ &\\
\citet{2008MNRAS.388.1487L} & $-$& $z<2$ for LLGRB excess& No &\\ 
 & $-$& $z>3$ for HLGRB excess  & &\\ \citet{2010ApJ...711..495B} & $z<1$ & $z>1$ & No & \\ \citet{2010MNRAS.406.1944W} & $z<3$ & $z>3$ & No  &\\ 
\citet{2019MNRAS.488.5823L} & $-$ & $z<1$,\ $2<z<7$ & Yes, $(1+z)^{3.5}$ & \\ 
\citet{2021ApJ...908...83T} & $z>1$ & $z<1$ & Yes, $(1+z)^{1.2}$ &\\
\citet{2022MNRAS.513.1078D} & $z>2$ & $z<1$ & Yes, $(1+z)^{2.88-3.92}$&  \\
\hline
\citet{2008ApJ...673L.119K} & $-$& $z<2$ for LLGRB excess& No,might cause & \multirow{7}{8em}
    {\textbf{Binary merger contamination}
    } 
\\ 
 & & $z>3$ for HLGRB excess & excess at low-$z$  & \\\citet{2008MNRAS.388.1487L} &$-$& $z<2$ for LLGRB excess& No & \\ 
 & & $z>3$ for HLGRB excess&  & \\ 
 \citet{2015ApJS..218...13Y} & $z>2$ & $z<2$ & Yes & \\ 
\citet{2019MNRAS.488.5823L} & $-$ & $z<1$,\ $2<z<7$ & Yes, $(1+z)^{3.5}$&  \\ 
\citet{2022MNRAS.513.1078D} & $z>2$ & $z<1$ & Yes, $(1+z)^{2.88-3.92}$ & \\
\hline
\citet{2007JCAP...07..003G} & $-$&Paucity:$1<z<2$  & $-$& \multirow{5}{8em}{\textbf{Low-$z$ excess due to LLGRB population}} \\ 
\citet{2008MNRAS.388.1487L} & $-$& $z<2$ for LLGRB excess & No&  \\ 
 & & $z>3$ for HLGRB excess & & \\ 
\citet{2015ApJS..218...13Y} & $z>2$ & $z<2$ & Yes&  \\
\citet{2015ApJ...806...44P} & $z>2$ & $z<2$ & Yes, $(1+z)^{2.3}$& \\\hline
\citet{2007JCAP...07..003G}
 &$-$&Paucity:$1<z<2$  & $-$  &\multirow{3}{8em}{\textbf{Selection effects}}\\
\citet{2010MNRAS.406.1944W} & $z<3$ & $z>3$ & No & \\ 
\citet{2015ApJS..218...13Y} & $z>2$ & $z<2$ & Yes& \\
\citet{2015ApJ...806...44P} & $z>2$ & $z<2$ & Yes, $(1+z)^{2.3}$ &\multirow{2}{8em}{\textbf{Selection effects}}  \\ 
\citet{2021ApJ...908...83T} & $z>1$ & $z<1$ & Yes, $(1+z)^{1.2}$& \\
\hline
\citet{Pescalli:2015yva} & Matches well & $z<2$ for & Yes, $(1+z)^{2.5}$ &\multirow{3}{8em}{\textbf{Incomplete sample/ differences in method}} \\
 & with SFR& simulated case & &\\
 \citet{2019MNRAS.488.5823L} & $-$ & $z<1$,\ $2<z<7$ & Yes, $(1+z)^{3.5}$& \\
 \hline
\citet{2004ApJ...609..935Y} & $z<2$ & $z>2$ & Yes, $(1+z)^{2.6}$&\multirow{2}{8em}{\textbf{$\theta\mathbf{_j}$ decreases/ $\mathbf{L}$ increases with $\mathbf{z}$}} \\
& & & & \\
\hline
\citet{2008ApJ...673L.119K} & $-$& $z<2$ for LLGRB excess & No,might cause & \multirow{2}{8em}{\textbf{Evolving IMF}}\\
& $-$& $z>3$ for HLGRB excess & excess at low-$z$& \\
%\hline
%\citet{2012MNRAS.423.2627W} & Across z & $-$ & Yes, (1+z)$^{2.3^{+0.56}_{-0.51}}$&\multirow{3}{14em}{Larger sample with z needed
%} \\ 
\hline
\citet{2019MNRAS.488.5823L} & $-$ & $z<1$,\ $2<z<7$ & Yes, $(1+z)^{3.5}$ & \multirow{3}{9em}{SFR peaks at low-$z$ for \textbf{lower mass galaxies}, giving high LGRB-RD \citep{2005ApJ...619L.135J}} \\
& & & & \\
& & & & \\
& & & & \\
\enddata
\tablecomments{Column 1: References, Column 2: $z$ range at which LGRB-RD matches with SFR, Column 3: $z$ range at which LGRB-RD does not match with SFR, Column 4: Luminosity evolution factor and Column 5: Solution proposed in references to explain the mismatch}
\end{deluxetable*}
\subsection{Selection criteria for limiting flux} \label{Limiting flux}

\begin{figure*}[!t]   
   \includegraphics[width=\textwidth]{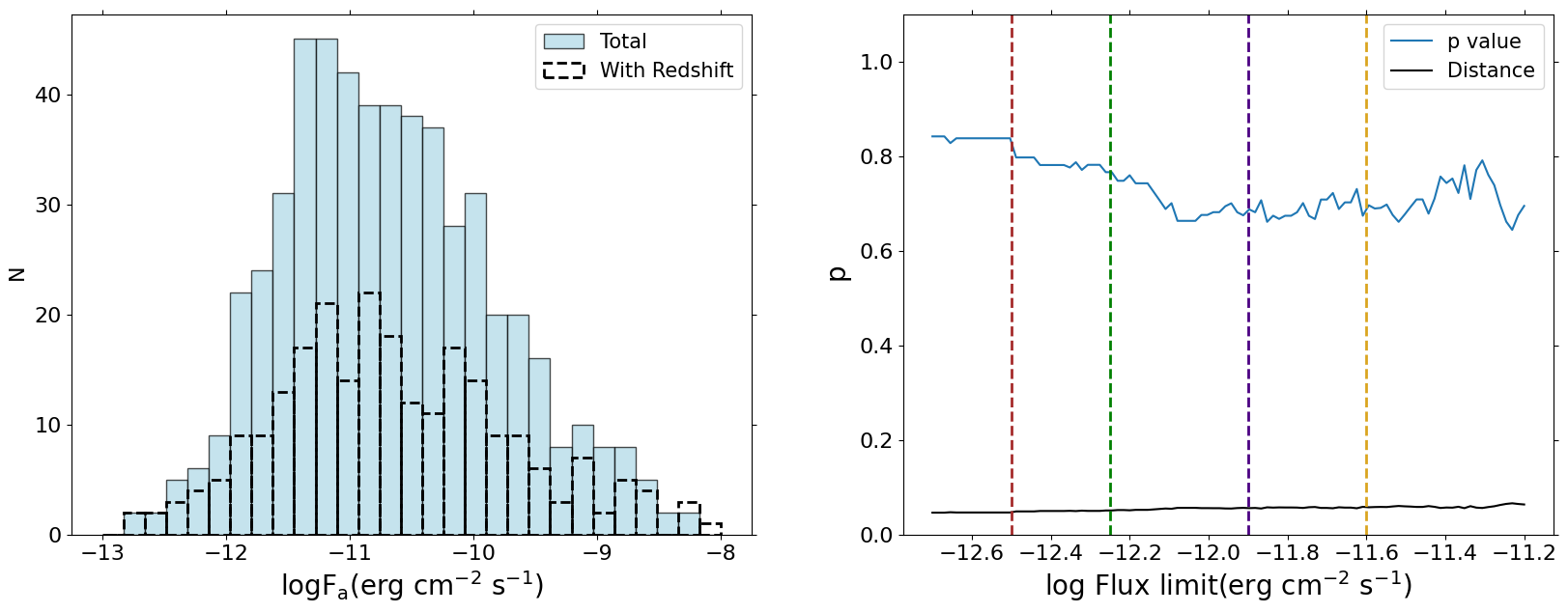} 
    \caption{Left panel: $\mathrm{log F_a (erg \ cm^{-2} \ s^{-1})}$ histogram for LGRBs with X-ray plateaus. The blue histogram bars correspond to LGRBs with and without redshift measurements. The black dashed histogram bars are for LGRBs with redshift measurements only. Right panel: p-value (blue line) and distance between the two histograms' cumulative distributions (black line) for varying Flux limit cuts. The red, green, purple and yellow dashed lines show the different $\mathrm{F_{lim} = -12.5, -12.25, -11.9}$ and $-11.6$, respectively. The chosen $\mathrm{F_{lim}}$ value of $-11.9$ is the purple dashed line.}
    \label{fig:Histogram_KStest}
\end{figure*}

\begin{figure*}[htbp]
    \centering
    \includegraphics[width=0.45\textwidth]{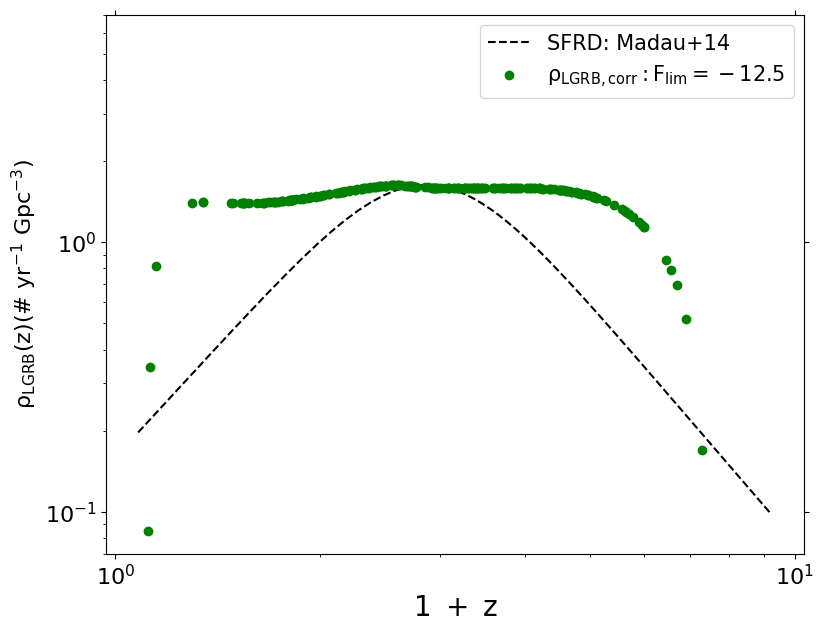}
    \includegraphics[width=0.45\textwidth]{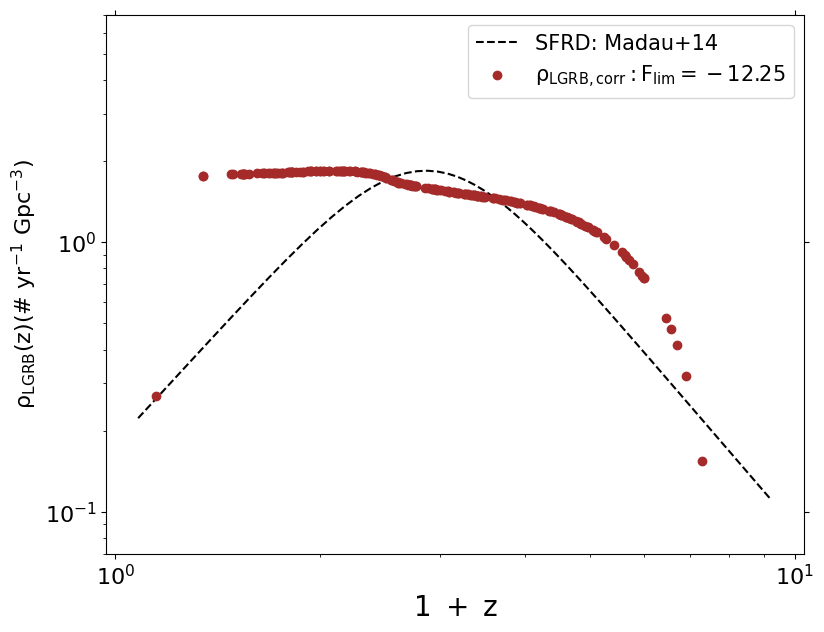}
    \includegraphics[width=0.45\textwidth]{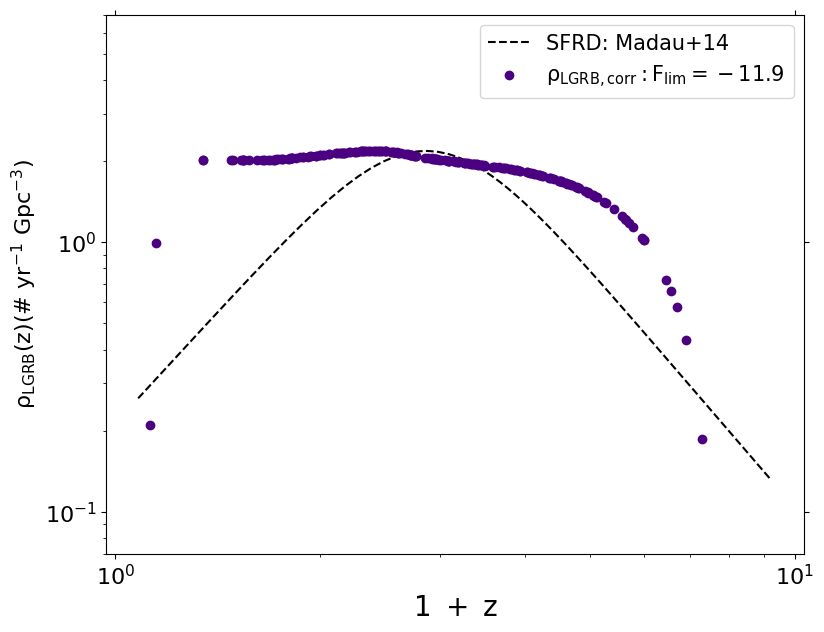}
    \includegraphics[width=0.45\textwidth]{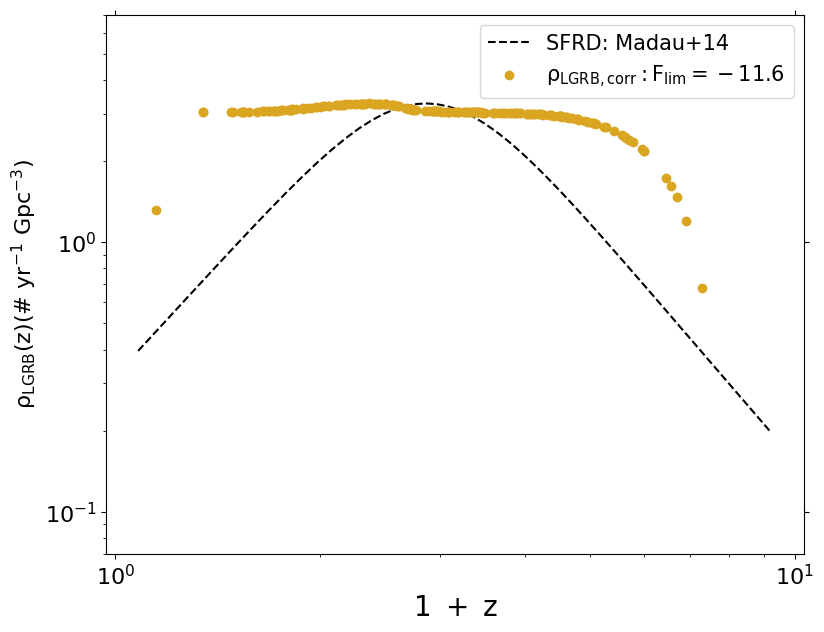}
    \caption{$\rho\mathrm{_{LGRB, corr}(z)}$ for four flux threshold cases of -12.5, -12.25, -11.9, and -11.6. The filled circles indicate the LGRB-RD corrected for intrinsic luminosity evolution. The filled circles are color coded with respect to the vertical dashed lines in the right panel of Figure \ref{fig:Histogram_KStest}. The black dashed lines show the SFRD from \citet{2014ARA&A..52..415M} study. The peak of SFRD is renormalized to peak of LGRB-RD for easier comparison.}
    \label{fig:rho_Flim}
\end{figure*}

We follow the approach mentioned in \citet{2021ApJ...914L..40D} to select the flux threshold value. This approach conducts the Kolmogorov-Smirnov (KS) test between two samples to test the completeness of the sample and exclude GRBs due to the detector sensitivity. Sample 1 includes all LGRBs with X-ray plateaus, and sample 2 includes only LGRBs with redshift measurements and X-ray plateaus. The histogram of these two samples for $\mathrm{log F_a}$ is shown in the left panel of Figure \ref{fig:Histogram_KStest}. Both samples are cut at different $\mathrm{log F_a}$ values to study the KS test statistics with varying flux limit. The behavior of the p-value and the distance between the empirical cumulative distributions for varying flux limit is shown in the right panel of Figure \ref{fig:Histogram_KStest}.  We test four flux limit cases shown by the vertical dashed lines to understand the effect of flux threshold on LGRB-RD. These cases correspond to logarithmic flux limit values of -12.5, -12.25, -11.9, and -11.6. They exclude 2\%, 5\%, 10\% and 15\%  GRBs from the sample with redshift measurements respectively. The 2\% case is due to the maximum flux limit for which the p-value is high, indicating the sample is still complete from the KS test. The remaining cases (5\%, 10\%, and 15\%) are explored since the p-value is high (p $>$ 0.68) to check the consistency of the rate density results between the four cases. Figure \ref{fig:rho_Flim} highlights that the nature of the LGRB-RD curve between the different flux threshold values is consistent. Among 5\%, 10\% and 15\% cases, we select the 10\% case as the most optimal choice as it has sufficient data points to describe the low-$z$ regime. 

%%%%%%%%%%%%%%%%%%%%%%%%%%%%%%%%%%%%%%%%%%%%%%
We do not select that flux threshold value for the $2\%$ case as negligible GRBs are excluded, i.e., essentially no detector limit. All results in the above sections are presented with a value of $\mathrm{F_{lim} = -11.9}$.

\end{document}